**Robustness and Transferability of Pix2Geomodel for Bidirectional Facies–Property Translation in a Complex Reservoir**

Abdulrahman Al-Fakih[1,2*], Nabil Sariah[1], Ardiansyah Koeshidayatullah[1,2], Sherif Hanafy[1,2] and SanLinn I. Kaka[1,2]

[1]Department of Geosciences, King Fahd University of Petroleum and Minerals, Dhahran, 31261, Saudi Arabia.

[2]Center for Integrative Petroleum Research, King Fahd University of Petroleum and Minerals, Dhahran, 31261, Saudi Arabia

Corresponding author: Abdulrahman Al-Fakih (alfakihabdulrahman2030@gmail.com)

## Abstract

Reservoir geomodeling plays a central role in subsurface characterization, but it remains challenging because conditioning data are sparse, geological heterogeneity is strong, and conventional geostatistical workflows often struggle to represent complex non-linear relationships between facies and petrophysical properties. This study evaluates the robustness and transferability of the Pix2Geomodel framework on a different and more geologically complex reservoir dataset with reduced vertical support. Unlike the original application, the present case includes a more heterogeneous reservoir-quality classification and only 54 retained layers, providing a stricter test of whether Pix2Pix-based image-to-image translation can preserve facies–property relationships under constrained data conditions. Facies, porosity, permeability, and clay volume (VCL) were extracted from a reference reservoir model, exported as aligned two-dimensional slices, augmented using consistent geometric transformations, and assembled into paired image datasets. Six bidirectional translation tasks were evaluated: facies→porosity, facies→permeability, facies→VCL, porosity→facies, permeability→facies, and VCL→facies. The Pix2Pix model, consisting of a U-Net generator and PatchGAN discriminator, was trained and evaluated using image-based metrics, visual comparison, and variogram-based spatial-continuity validation. Results show that the model preserves the dominant geological architecture and main spatial-continuity trends across the evaluated tasks, with strongest performance observed for facies–porosity and VCL–facies translation. Quantitatively, facies→porosity achieved the highest pixel accuracy and frequency-weighted intersection over union values of 0.9326 and 0.8807, respectively, while VCL→facies achieved the highest mean pixel accuracy and mean intersection over union values of 0.8506 and 0.7049, respectively. Permeability-related tasks remain more challenging because permeability has a wider dynamic range and is controlled by pore connectivity, diagenesis, and sub-facies heterogeneity in addition to facies architecture. These findings demonstrate that Pix2Geomodel can be transferred beyond its original case study as a practical, data-efficient framework for rapid bidirectional facies–property translation, with implications for both local-scale heterogeneity assessment and larger-scale reservoir modeling workflows.

## 1. Introduction

Reservoir geomodeling is a cornerstone of subsurface characterization, supporting applications such as hydrocarbon production, groundwater management, and carbon storage (Cannon et al., 2024; Eltom et al., 2020; Isah et al., 2025; Kazemi et al., 2025; Saraih, et al., 2023). Its primary objective is to represent the spatial distribution of geological structures, facies, and petrophysical properties in a way that reliably predicts subsurface behavior. Despite its importance, reservoir modeling remains a persistent challenge because subsurface datasets/observations are sparse, geological heterogeneity is strong, and conventional workflows rely heavily on geostatistical methods and expert interpretation, which, although widely used, often struggle to capture complex, non-linear spatial relationships (Khogali et al., 2022; Rajput & Pathak, 2025). This limitation becomes more pronounced in structurally and sedimentologically complex reservoirs, where uncertainty in inter-well regions can significantly impact model reliability (Jolley & Knipe, 2007; Liu et al., 2026; Mikes et al., 2001).

Beyond depositional facies, reservoir performance is often controlled by fine-scale variations in pore geometry, connectivity, transmissibility, fracture, and shale content. While facies models provide an essential framework for describing subsurface heterogeneity, they alone are insufficient for decision-making. Critical reservoir behavior emerges from the interaction between facies and petrophysical properties such as porosity, permeability, and clay volume (VCL). Accurately capturing these facies–property relationships therefore remain a central challenge in reservoir modeling (Khogali et al., 2022; Mikes et al., 2001; Saraih, et al., 2023, 2024).

Recent advances in subsurface generative modeling have significantly expanded the capabilities of reservoir characterization. Early developments focused on unconditional facies generation, demonstrating that GAN-based methods can reproduce complex geological patterns while honoring conditioning data (Banh & Strobel, 2023; Hadid et al., 2024; Hammouri et al., 2026; Lv, 2023; Olowookere et al., 2025). This was followed by the adoption of U-Net and Pix2Pix-style architectures, which established image-to-image translation as an effective framework for conditional facies simulation (Al-Fakih, et al., 2026; Isola & Efros, 2017; Saraih & Kaka, 2026; Zeedan, & Abushaikha, 2025; Zhang & Azevedo, 2021). Subsequent improvements enhanced multi-facies realism through techniques such as one-hot encoding, hybrid discriminators, refined loss functions, and multiscale conditioning strategies, enabling better preservation of depositional structures and spatial variability (Chan & Elsheikh, 2019; Dupont et al., 2018; Liu et al., 2026; Misra et al., 2024; Song & Mukerji, 2026).

The scope of generative modeling has since broadened beyond facies generation toward conditioned geomodeling, inverse modeling, and cross-domain translation (Al-Fakih et al., 2026; Feng & Mosegaard, 2025; Song & Hou, 2021). In particular, generative networks have been integrated with inversion workflows and training-image-based geostatistics, where conditional GANs serve as efficient priors for stochastic facies inversion(Feng et al., 2025; Merzoug & Pyrcz, 2025; Song & Hou, 2021; Song & Lyu, 2022). Similarly, training-image-based generative approaches have been combined with gradual deformation techniques to embed learned geological patterns within uncertainty quantification frameworks (Al-Fakih & Kaka, 2026; Garayt et al., 2025; Song & Feng, 2026). These developments highlight a shift from simple image synthesis toward conditional sampling and model updating.

Despite this progress, most existing workflows remain focused on facies generation or conditioning. Limited attention has been given to bidirectional translation between facies and petrophysical properties, which is essential for integrated reservoir modeling (Feng & Hansen, 2024; Garayt et al., 2025; Zhang et al., 2024). Recent advances in generative artificial intelligence (GAI) provide a promising direction by enabling models to learn spatial relationships directly from geological and petrophysical data, rather than relying solely on predefined statistical assumptions (Dong et al., 2026; Feng & Hansen, 2024; Sun & Arnold, 2023; Yang, & Wang, 2025; Zhang et al., 2024).

Within this context, Pix2Geomodel was introduced as a Pix2Pix-based conditional GAN workflow for facies–property translation (Al-Fakih et al., 2026). The original study was conducted on a relatively simple reservoir setting, characterized by a limited facies architecture and a large number of vertical slices (exceeding 280 layers). The results demonstrated that image-to-image learning can reproduce reservoir facies and petrophysical properties while preserving geological realism through quantitative evaluation metrics and variogram-based spatial validation. However, the original workflow was tested under a more favorable data setting, with greater vertical support and simpler facies architecture than many practical reservoir-modeling problems. Its robustness under a more heterogeneous reservoir setting, reduced vertical support, and expanded bidirectional facies–property tasks therefore remain insufficiently evaluated.

The present study addresses this gap by applying the Pix2Geomodel workflow to a different and more challenging reservoir dataset. The new case includes a seven-class reservoir-quality representation, stronger heterogeneity in petrophysical properties, and a simplified vertical framework of retained layers. Rather than proposing a new generative architecture, this work evaluates whether the existing Pix2Pix-based workflow remains transferable and geologically reliable when applied under more constrained conditions. The study also documents practical implementation constraints, including data alignment, augmentation consistency, image resizing, computational requirements, and validation challenges, which are essential for applying the workflow beyond the original case study. The main contributions of this study are: (1) evaluation of Pix2Geomodel transferability on a new and more complex reservoir dataset; (2) implementation of six bidirectional facies–property translation tasks involving porosity, permeability, and VCL;

(3) assessment of model behavior across high-agreement, typical, and challenging examples; (4) validation using both image-space metrics and spatial-continuity analysis; and (5) identification of workflow constraints and future development directions, including attention-enhanced, multiscale, diffusion-based, and three-dimensional extensions that are proposed for subsequent studies.

## 2. Methodology

### 2.1 Data preparation and pairing

The workflow starts from a reference three-dimensional reservoir model containing facies, porosity, permeability, and VCL (Fig. 1). The original grid consists of 365 × 248 × 55 cells, from which 54 layers were retained for modeling (Table 1). These models represent the key geological and petrophysical properties required to describe reservoir behavior, and they were extracted in a consistent manner to preserve spatial alignment across all variables.

The facies model was classified into seven reservoir-quality classes—tight, very poor sand, poor sand, fair sand, good sand, good clean sand, and very good sand—based on effective porosity (PHIE) and VCL. This classification reflects variations in rock quality and depositional characteristics within the reservoir. In addition, porosity, permeability, and VCL were derived from core data, well logs, and correlation.

The facies model is represented using these seven reservoir-quality classes, while the corresponding petrophysical domains, porosity, permeability, and VCL, exhibit distinct spatial patterns and variability (Fig. 1). Porosity generally shows relatively smooth spatial trends governed by facies architecture, whereas permeability displays higher heterogeneity and a wider dynamic range due to its sensitivity to pore structure and connectivity. VCL exhibits intermediate behavior, capturing shale distribution and its influence on reservoir quality (Fig. 7). Together, these domains highlight the complexity of the geological system and the challenge of learning consistent cross-domain relationships.

All models were then exported as high-resolution PNG images using fixed, variable-specific legends to ensure pixel-wise spatial correspondence (Fig. 2). Porosity and VCL were exported using fixed legend-controlled RGB gradients over ranges of 0.00–0.225 and 0.00–0.50, respectively, while permeability was exported using a logarithmic color scale over 0.01–1000 mD due to its wider dynamic range (Fig. 1). Each single-property slice had a resolution of 8184 × 12045 pixels. These aligned domains serve as the foundation for paired-slice generation and subsequent bidirectional facies–property translation. Table 1 summarizes the extraction and dataset settings, and Table 2 lists the encoding scheme for each domain.

Table 1. Summary of reservoir-slice extraction, data augmentation, and paired-image generation used for training, validation, and testing.

| Item | Description |
|---|---|
| Reference grid size | 365 × 248 × 55 |
| Retained layers for modeling | 54 |
| Exported domains | Facies, porosity, permeability, VCL |
| Export image size | 8184 × 12045 px |
| Original images per domain | 54 |
| Augmentation per retained layer | 30 |
| Augmentation operations | ±10° rotation, 10% shift, 10% x/y shear, ±10% zoom, optional horizontal flip |
| Augmented images per domain | 1620 |
| Total paired images per task | 1674 |
| Dataset split | 70% training, 15% validation, 15% testing |
| Images per split (original + augmented) | 1171 training, 251 validation images, 252 testing |
| Training image size | 2048 × 1024 px |

Table 2. Encoding strategy and export settings for the facies and petrophysical image domains used in paired image-to-image translation.

| Domain | Data type | Encoding | Range / classes | Notes |
|---|---|---|---|---|
| Facies | Categorical | Seven-class categorical legend | tight; very poor sand; poor sand; fair sand; good sand; good clean sand; very good sand | Palette PNG; background index 255 |
| Porosity | Continuous | Legend-controlled RGB gradient | 0.00–0.225 | Palette PNG; fixed export legend |
| Permeability | Continuous | Log-scaled RGB gradient | 0.01–1000 mD | Palette PNG; logarithmic display scale |
| VCL | Continuous | Legend-controlled RGB gradient | 0.00–0.50 | Palette PNG; fixed export legend |

To enlarge the effective dataset while preserving pixel alignment, a consistent augmentation strategy was applied to all domains. The same augmentation policy was used independently for each property and then matched by layer number and augmentation index to maintain correspondence between input and target images. The augmentation operations included random rotations (±10°), translations (up to 10% of image width or height), x- and y-shear (up to 0.10), zoom variation (±10%), and optional horizontal flipping. For each retained layer, thirty augmented realizations were generated, resulting in 1620 augmented images per domain in addition to the 54 original slices. The original and augmented paired images were divided into training, validation, and testing subsets using a 70/15/15 split. The reported metrics are therefore interpreted as image-space performance indicators rather than strict unseen-layer generalization metrics.

For each task, spatially corresponding source and target slices were concatenated horizontally to form paired images. Given the single-slice size of 8184 x 12045 px, horizontally concatenated paired images had a raw size of 16368 x 12045 px. Six translation tasks were defined: facies-to-porosity, facies-to-permeability, and facies-to-VCL, together with three inverse tasks: porosity-to-facies, permeability-to-facies, and VCL-to-facies. The final dataset for each task contained 1674 paired samples (54 original + 1620 augmented), which were split into training (70%), validation (15%), and testing (15%) subsets, corresponding to 1171, 251, and 252 images, respectively. The paired datasets were used to train the Pix2Pix model, a conditional generative adversarial network (cGAN), followed by evaluation on held-out test data. This workflow enables consistent mapping between geological and petrophysical domains while preserving spatial coherence.

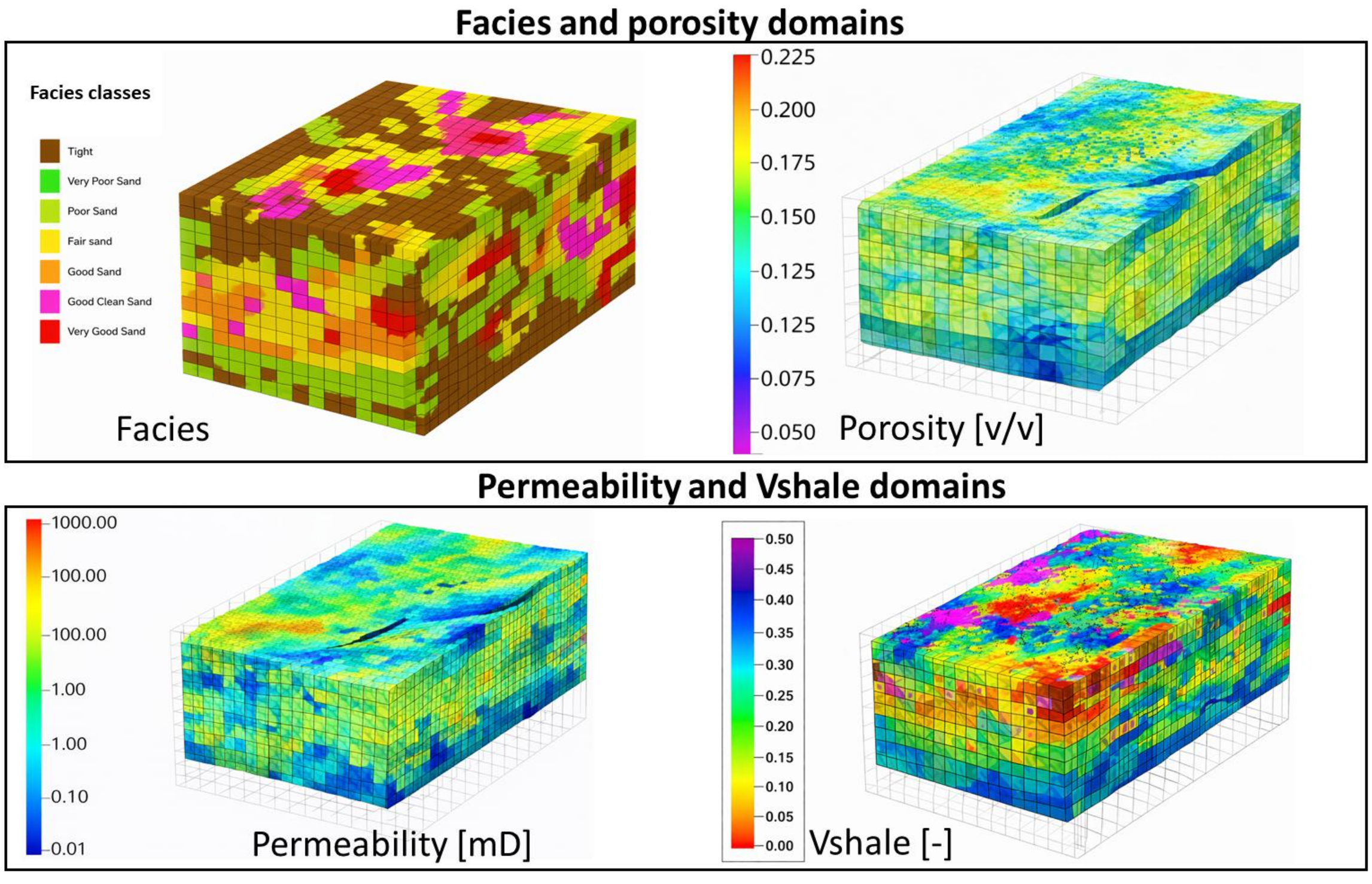

Figure 1. Representative 3D facies and petrophysical-property domains from the reference reservoir model. The model includes facies, porosity, permeability, and VCL distributions, from which aligned two-dimensional slices were extracted and exported for paired image generation and subsequent Pix2Pix training.

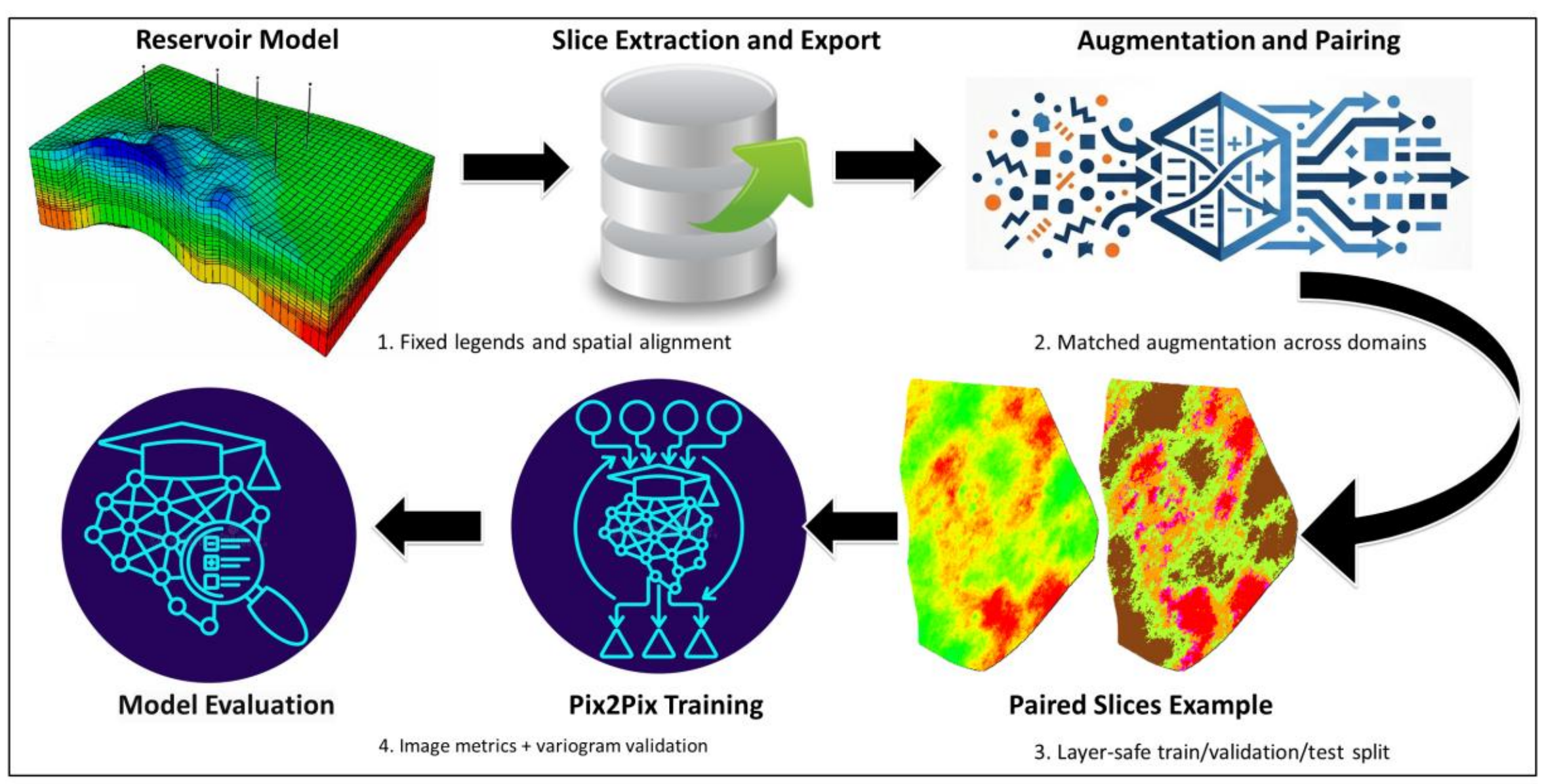

Figure 2. Dataset generation and utilization workflow for the extended Pix2Geomodel implementation. Reservoir property models were extracted from the reference 3D geomodel and exported as aligned 2D slices using fixed variable-specific legends. The slices were augmented using matched transformations and concatenated into paired input–target datasets for supervised Pix2Pix training. Compared with the original Pix2Geomodel workflow, the present implementation emphasizes high-resolution paired-image handling, consistent augmentation, bidirectional facies–property task design, and spatial-continuity validation under reduced vertical support.

## 2.2 Pix2Pix architecture and training configuration

The modeling framework is based on the standard Pix2Pix of cGAN, which learns a supervised mapping between an input image domain and a target image domain using the paired datasets described in Section 2.1. For training, each paired RGB image was resized from the original resolution of 16368 x 12045 px to 2048 × 1024 px and normalized to the range [−1, 1] to ensure stable optimization and maintain consistent scaling across domains.

The generator follows a canonical U-Net architecture composed of eight downsampling blocks and seven upsampling blocks connected through skip connections (Fig. 3). This encoder–decoder structure enables the model to capture both large-scale contextual features and fine-scale spatial details, which is essential for preserving geological continuity and structural patterns during translation. The skip connections allow high-resolution features to be directly transferred from the encoder to the decoder, improving reconstruction fidelity and reducing information loss (Fig. 3).

The discriminator adopts a PatchGAN architecture, which evaluates the realism of local image patches rather than the entire image. This design encourages the generator to produce outputs that are locally consistent and geologically realistic. The model is trained using a combination of adversarial loss and L1 reconstruction loss, where the adversarial term promotes realism and the L1 term enforces similarity to the reference image. The L1 loss is weighted by λ = 100 to balance global structure preservation with local detail accuracy.

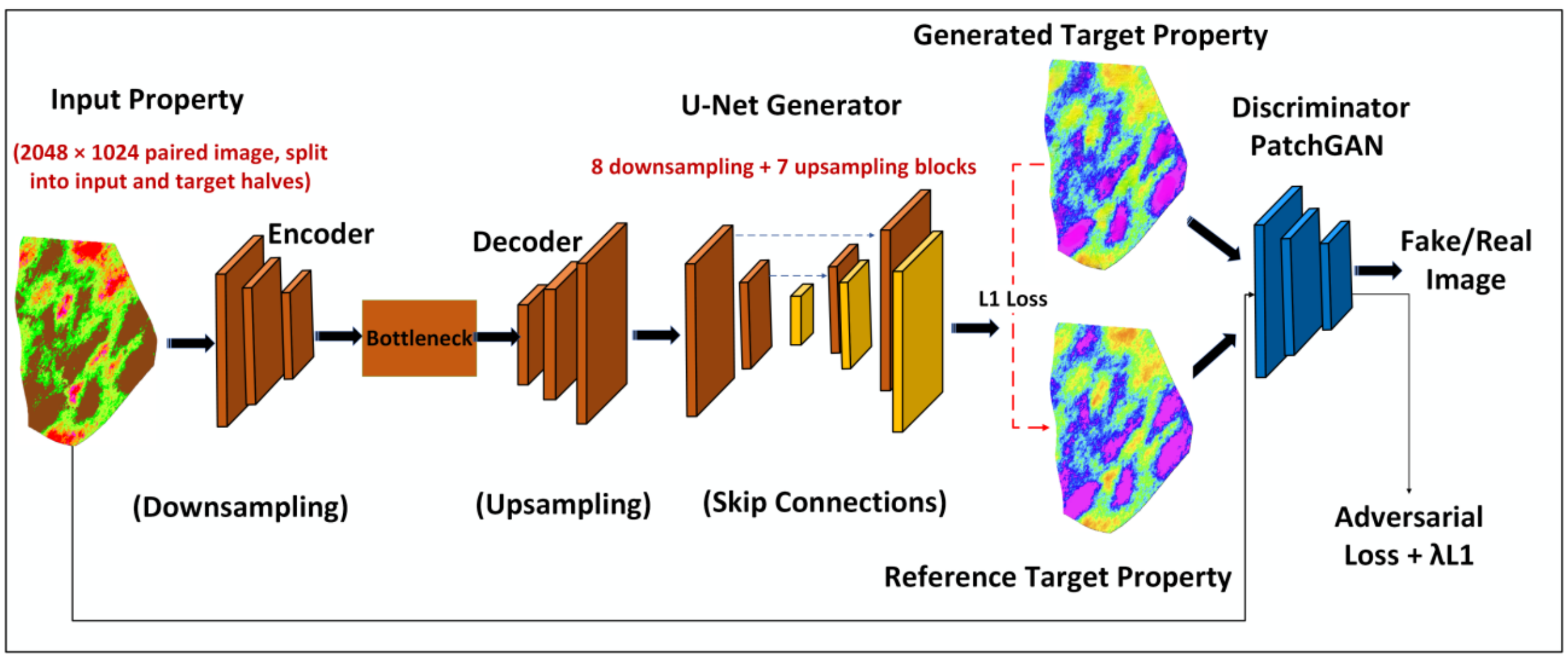

Figure 3. Pix2Pix architecture used in the current Pix2Geomodel extension. The framework uses a U-Net generator with eight downsampling blocks, seven upsampling blocks, and skip connections to translate an input reservoir

image into the target domain. A PatchGAN discriminator evaluates local patch realism by comparing generated and reference target images, while training is guided by adversarial loss and L1 reconstruction loss with $\lambda = 100$.

Both the generator and discriminator were optimized using the Adam optimizer with a learning rate of 2 × 10^−4 and β1 = 0.5. Training was performed with a batch size of 1 and a shuffle buffer size of 400, consistent with standard Pix2Pix implementations for high-resolution image translation. Additional online augmentation was applied during training, including padding, approximately 12% upscaling, random cropping back to the target input size, and random horizontal flipping. These augmentations improve model generalization while preserving spatial relationships between input and target domains.

The evaluated translation tasks include facies→porosity, facies→permeability, facies→VCL, porosity→facies, permeability→facies, and VCL→facies. All six translation tasks were trained for 15,000 steps. The complete network architecture, preprocessing steps, and training configuration are summarized in Table 3.

Table 3. Pix2Pix architecture, preprocessing, training settings, and evaluated translation tasks used in the extended Pix2Geomodel workflow.

| Item | Setting |
|---|---|
| Model family | Pix2Pix conditional GAN |
| Generator | U-Net with 8 downsampling and 7 upsampling blocks with skip connections |
| Discriminator | PatchGAN |
| Input representation | Side-by-side paired RGB images |
| Raw paired image size | 16368 × 12045 px |
| Training image size | 2048 × 1024 px |
| Normalization | [−1, 1] |
| Online augmentation | Padding, ~12% upscale, random crop, horizontal flip |
| Loss function | Adversarial loss + L1 reconstruction loss |
| L1 weight (λ) | 100 |
| Optimizer | Adam |
| Learning rate | 2 × 10^−4 |
| β1 | 0.5 |
| Batch size | 1 |
| Shuffle buffer | 400 |
| Evaluated tasks | Facies→Porosity; Facies→Permeability; Facies→VCL; Porosity→Facies; Permeability→Facies; VCL→Facies |
| Training steps | 15,000 steps for all reported tasks |
| Compute resources | Single NVIDIA RTX A5500 GPU (24 GB VRAM), Approximately 30 min per task |

### 2.3 Evaluation protocol

Model outputs were assessed using a combination of quantitative image-based metrics and geological consistency. Quantitative evaluation included pixel accuracy (PA), mean pixel accuracy (mPA), mean intersection over union (mIoU), and frequency-weighted IoU (FWIoU), computed between the generated and reference images. These metrics are useful for monitoring model behavior during image-to-image translation, particularly for categorical facies predictions and consistency in color-encoded outputs.

However, since the objective extends beyond image similarity, geological validation was also performed. Qualitative comparisons between reference and generated slices were used to assess the reproduction of major facies distributions, structural trends, and cross-domain relationships. In addition, experimental variograms were used to examine whether the generated outputs retained the essential spatial continuity of the reference reservoir model.

This combined evaluation framework enables assessment from both machine-learning and geomodeling perspectives. The focus is therefore not only on visual agreement, but also on whether the learned facies–property relationships remain geologically coherent under increased facies complexity and reduced vertical resolution.

# 3. Results

## 3.1 Forward translation from facies to petrophysical properties

The forward translation results are evaluated on a test set of 252 images for each task. The predicted outputs reproduce the main spatial organization of the target petrophysical properties, with varying levels of agreement across the different properties. Visual comparisons of representative slices are shown in Figures 4–6.

For facies-to-porosity translation, the generated images reproduce the dominant spatial patterns observed in the reference slices (Fig. 4). Across the test dataset, many slices show strong agreement with the reference (Fig. 4A1-3) where both large-scale structures and local variations are closely matched. In other slices, the overall architecture is preserved, while local transitions appear smoother (Fig. 4B1-3). In a smaller number of examples, thin features are less sharply defined (Fig. 4C1-3), although the main spatial organization remains consistent. This task achieved PA = 0.9326, mPA = 0.6079, mIoU = 0.5594, and FWIoU = 0.8807 (Table 4).

For facies-to-VCL translation, the predicted images reproduce the main shale-rich and cleaner regions and maintain the dominant spatial continuity across the test slices (Fig. 5). The visual comparison shows consistent agreement at the scale of broad geological patterns (Fig. 5A1-3, 5B1-3). In some examples, transitions between adjacent VCL values appear smoother, particularly in regions with gradual changes (Fig. 5C1-3). This task achieved PA = 0.6608, mPA = 0.7608, mIoU = 0.5593, and FWIoU = 0.5078 (Table 4).

For facies-to-permeability translation, the generated outputs reproduce the main high- and low-permeability regions and preserve the overall spatial organization (Fig. 6). Across the test dataset, the large-scale connectivity is maintained in most slices (Fig. 6A1-3, 6B1-3). However, visual comparisons show greater variation at the local scale, where thin streaks and high-contrast features are less clearly resolved in some examples (Fig. 6C1–C3). This task achieved PA = 0.7813, mPA = 0.5694, mIoU = 0.4546, and FWIoU = 0.6530 (Table 4).

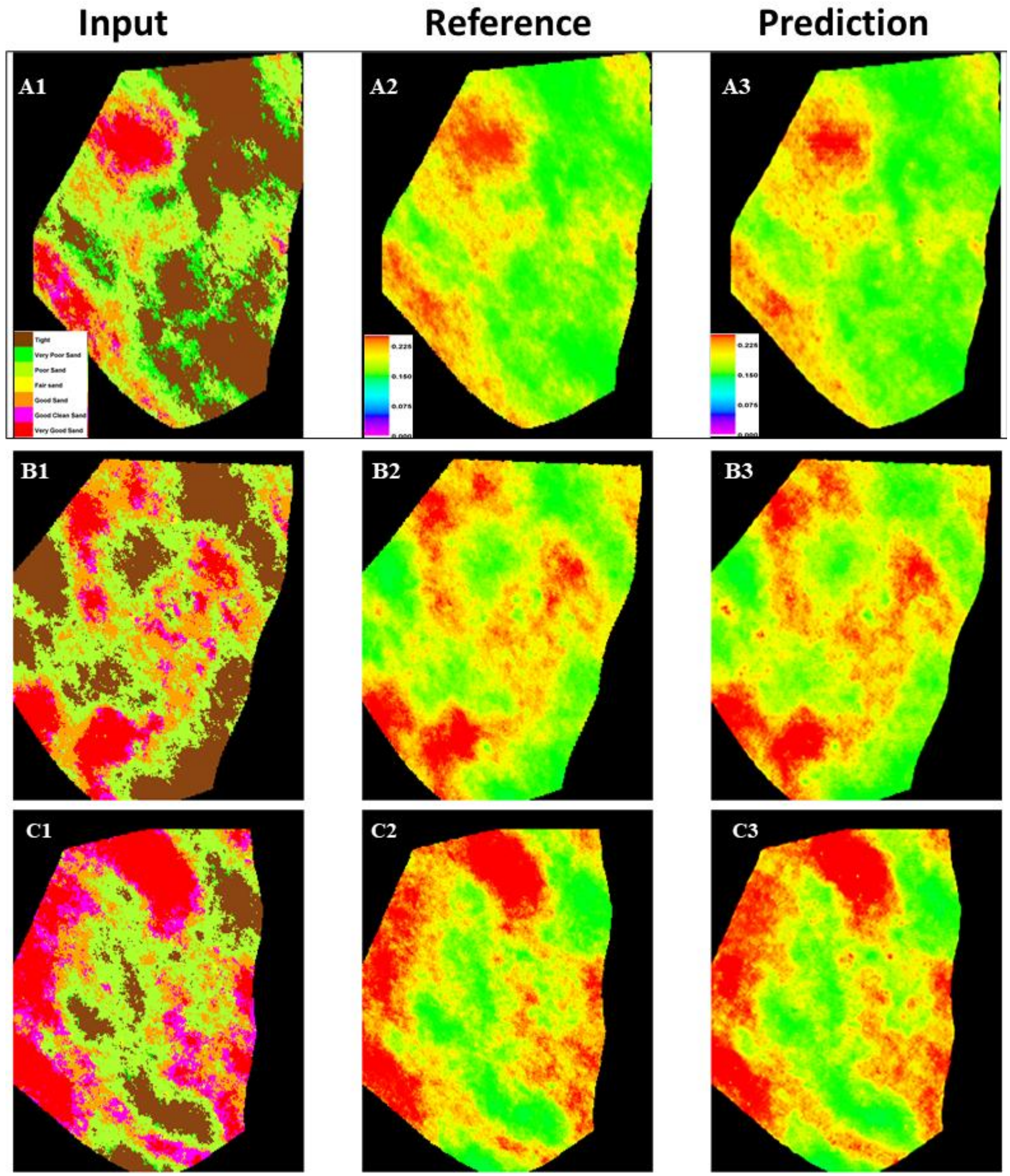

Figure 4. Facies-to-porosity translation for three representative test cases. Each row compares the input facies slice, the reference porosity slice, and the predicted porosity slice. A1-A3 illustrate a high-agreement case, B1-B3 a typical case, and C1-C3 a challenging case. The generated porosity fields preserve the dominant geological architecture and the main large-scale continuity trends, although local transitions become smoother in the more difficult example.

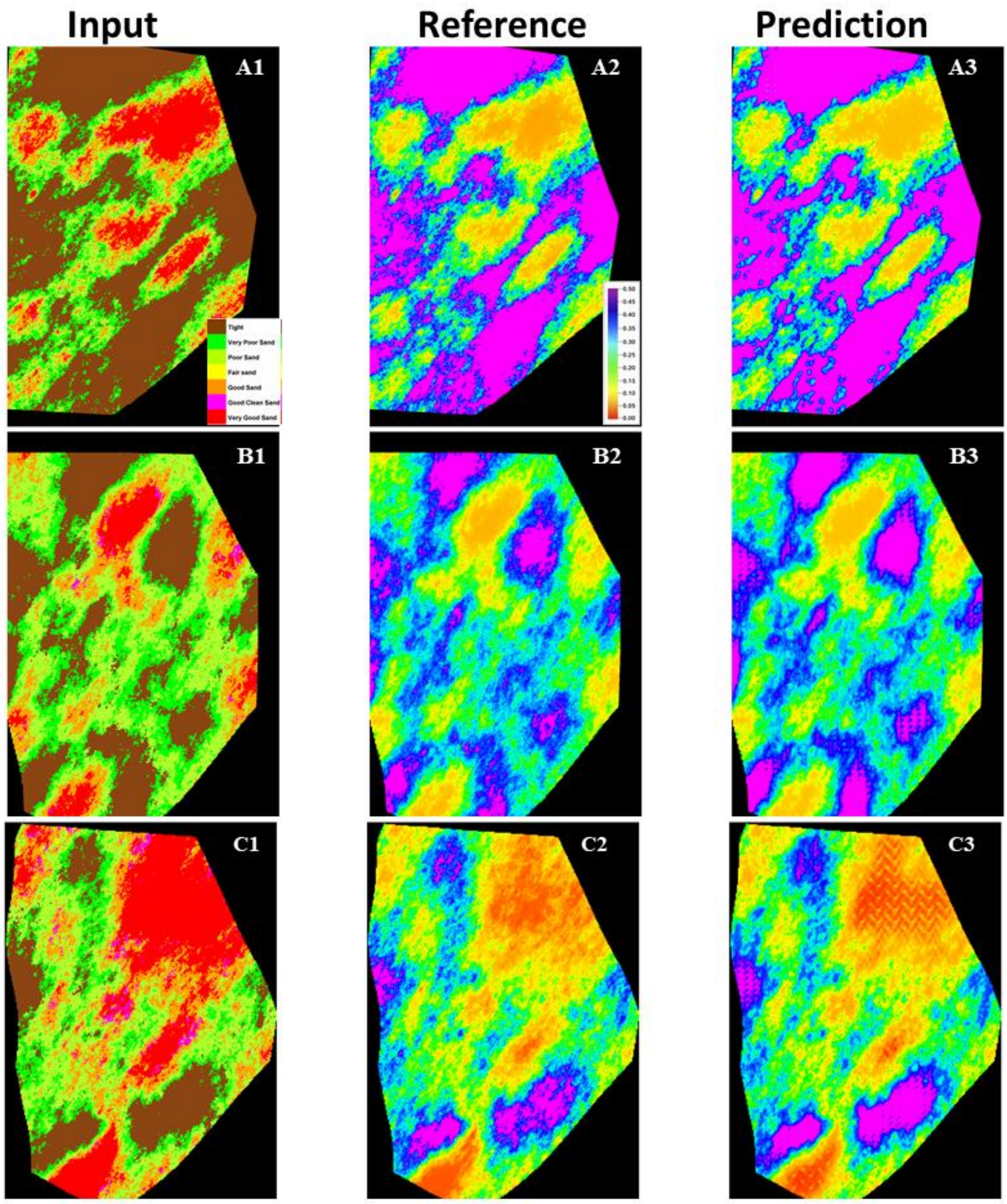

Figure 5. Facies-to-VCL translation for three representative test cases. Each row compares the input facies slice, the reference VCL slice, and the predicted VCL slice. From top to bottom, A1-A3 illustrate a high-agreement case, B1-B3 a typical case, and C1-C3 a challenging case. The predicted VCL maps reproduce the broad shale-rich and cleaner domains while showing increased smoothing in transitional zones.

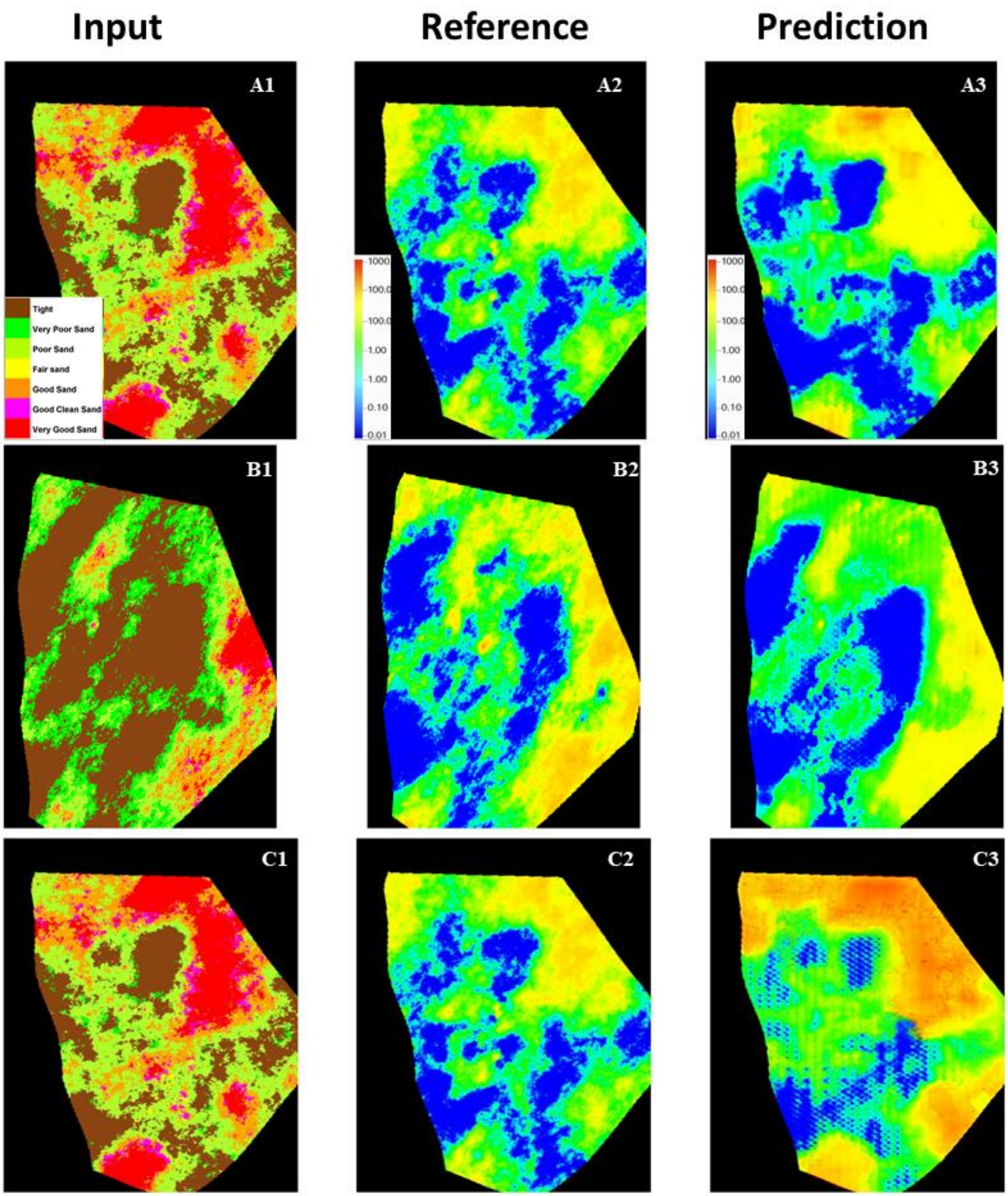

Figure 6. Facies-to-permeability translation for three representative test cases. Each row compares the input facies slice, the reference permeability slice, and the predicted permeability slice. A1-A3 illustrate a high-agreement case, B1-B3 a typical case, and C1-C3 a challenging case. The model captures the main connected permeability architecture but struggles more with thin streaks and extreme contrasts than in the porosity and VCL tasks.

### 3.2 Reverse translation from petrophysical properties to facies

The inverse translation results are evaluated on 252 test images per task using both quantitative metrics (PA, mPA, mIoU, and FWIoU) and visual comparison of representative slices (Fig. 7).

For porosity-to-facies translation, the predicted facies maps reproduce the main facies distribution and preserve the dominant spatial organization (Fig. 7A). Across the test dataset, many slices show consistent agreement with the reference in terms of large-scale facies belts. In some examples, local boundaries between adjacent classes appear smoother. This task achieved PA = 0.9060**,** mPA = 0.5898**,** mIoU = 0.5129**,** and FWIoU = 0.8418 (Table 4).

For VCL-to-facies translation, the generated facies maps reproduce the separation between cleaner and shalier regions and preserve the overall facies structure (Fig. 7B). Visual comparisons show consistent alignment of major facies zones across most test slices, with clearer separation between classes in many examples. This task achieved PA = 0.8440**,** mPA = 0.8506**,** mIoU = 0.7049**,** and FWIoU = 0.7429 (Table 4).

For permeability-to-facies translation, the predicted outputs reproduce the main facies architecture and preserve the dominant spatial trends (Fig. 7C). Across the test dataset, the large-scale structure is maintained, while local class boundaries show more variability and appear more diffuse in some slices. This task achieved PA = 0.7948**,** mPA = 0.4274**,** mIoU = 0.3500, and FWIoU = 0.6955 (Table 4). Values are image-space metrics computed from color-encoded outputs and should be interpreted as pattern-similarity indicators rather than direct petrophysical error measures.

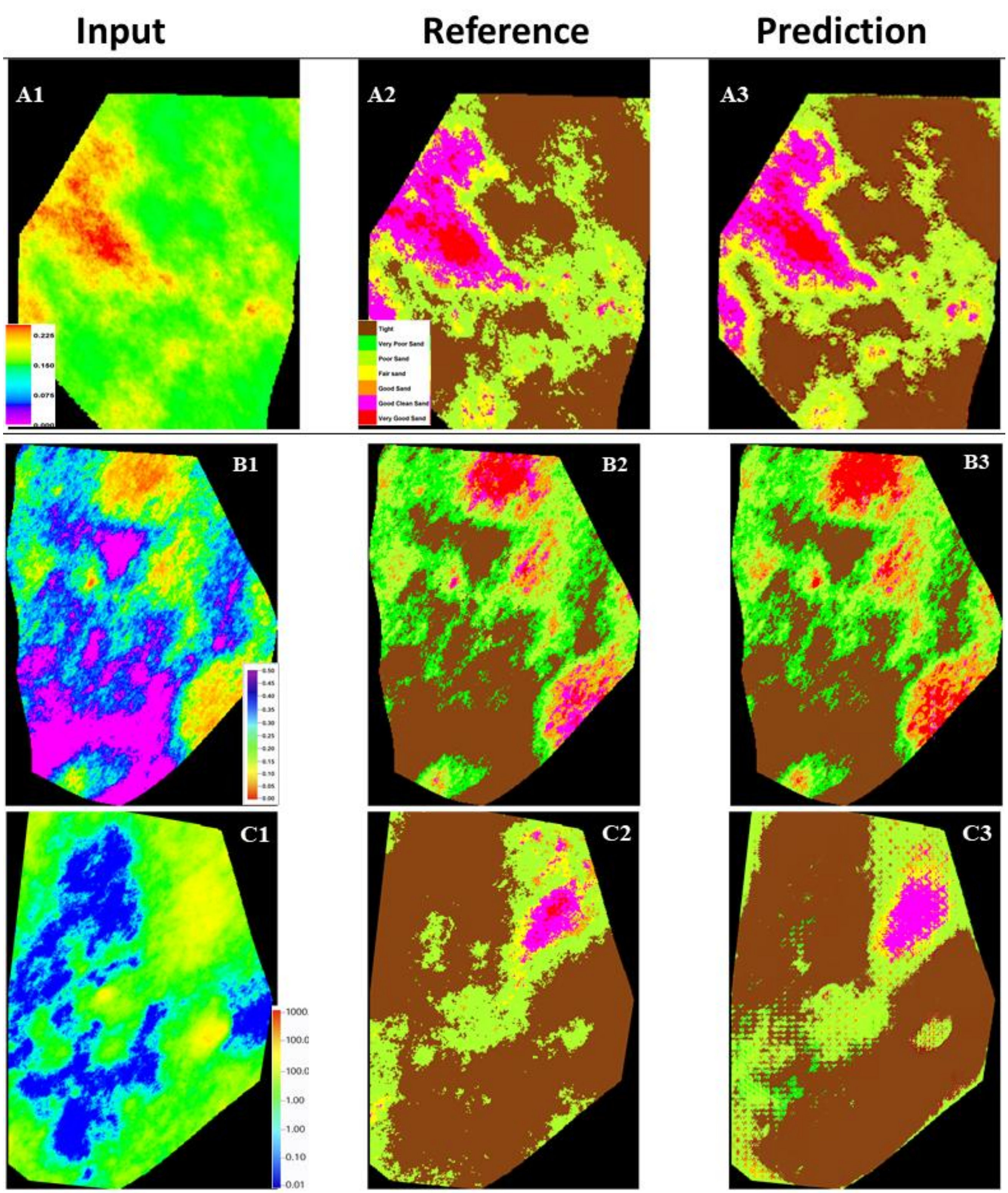

Figure 7. Inverse property-to-facies translation for representative test cases. From top to bottom, A1–A3 show porosity→facies, B1–B3 show VCL→facies, and C1–C3 show permeability→facies. In each row, the panels compare the input property slice, reference facies slice, and predicted facies slice. The VCL→facies case shows the clearest facies separation, whereas permeability→facies exhibit greater local class confusion and boundary smoothing.

### 3.3 Quantitative evaluation and spatial-continuity validation

Spatial-continuity validation was performed using experimental variograms computed for porosity, permeability, facies, and VCL, comparing generated outputs with the corresponding reference data (Fig. 8). For porosity, the generated variogram closely follows the reference curve across the full lag range (0–25 pixels), with both curves showing a gradual increase in semivariance and reaching similar sill values. The associated absolute-error curve remains low across all lags, with a mean absolute error (MAE) of 0.0006, indicating minimal deviation between generated and reference spatial structures.

For permeability, the generated variogram also follows the general trend of the reference curve, although a slight underestimation is observed at intermediate lags. The semivariance increases with lag in both cases, reaching comparable sill behavior, while the absolute-error curve shows higher values relative to other properties, with a peak around mid-range lags and an overall MAE of 0.0046.

For facies, the variogram comparison was performed using generated facies maps obtained from the inverse translation tasks (porosity→facies, permeability→facies, and VCL→facies) and compared against the original facies model. The generated and reference variograms show close agreement across the lag range, with both curves exhibiting a consistent increase in indicator semivariance and approaching similar sill values. The absolute-error curve shows moderate variability across lags, with an MAE of 0.0036.

For VCL, the generated variogram closely matches the reference curve, with both showing a rapid increase at short lags followed by a gradual stabilization toward the sill. The absolute-error values remain relatively low across the lag range, with MAE of 0.0029, indicating close correspondence between generated and reference spatial continuity.

Table 4. Quantitative evaluation of the six bidirectional facies–property translation tasks on the held-out test set. Values are image-space metrics computed from color-encoded outputs and should be interpreted as pattern-similarity indicators rather than direct petrophysical error measures. Bold values indicate the best performance for each metric.

| Task | Steps | Test images | PA | Mean PA | mIoU | FWIoU |
|---|---|---|---|---|---|---|
| Facies → Porosity | 15,000 | 252 | **0.9326** | 0.6079 | 0.5594 | **0.8807** |
| Facies → Permeability | 15,000 | 252 | 0.7813 | 0.5694 | 0.4546 | 0.6530 |
| Facies → VCL | 15,000 | 252 | 0.6608 | 0.7608 | 0.5593 | 0.5078 |
| Porosity → Facies | 15,000 | 252 | 0.9060 | 0.5898 | 0.5129 | 0.8418 |
| Permeability → Facies | 15,000 | 252 | 0.7948 | 0.4274 | 0.3500 | 0.6955 |
| VCL → Facies | 15,000 | 252 | 0.8440 | **0.8506** | **0.7049** | 0.7429 |

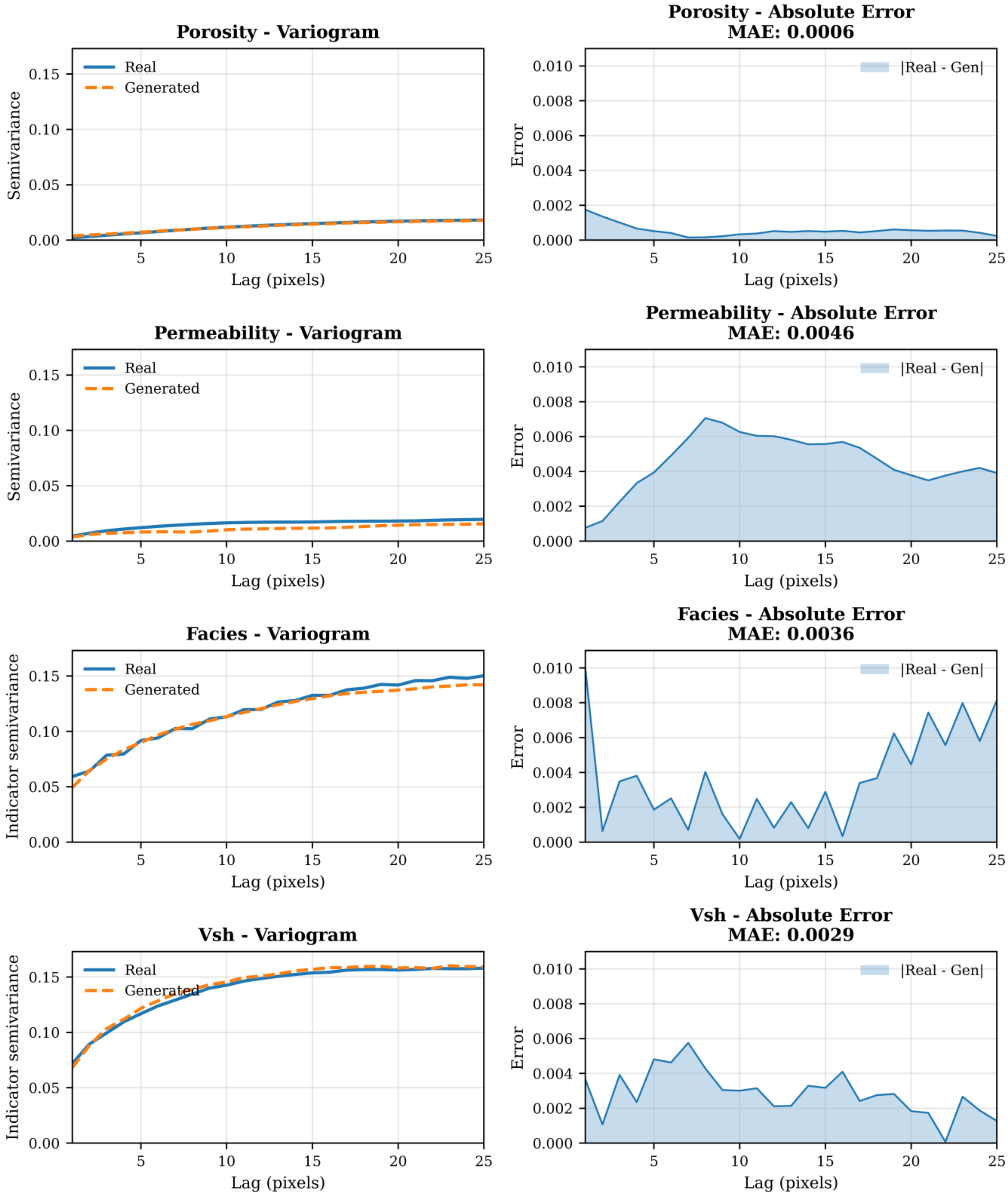

**Figure 8**. Spatial-continuity validation of the generated outputs. Experimental variograms for porosity, permeability, facies, and VCL are compared with the corresponding reference variograms, and absolute-error curves are reported for each domain. The close variogram agreement indicates that the generated images retain the dominant spatial continuity of the reference reservoir.

# 4. Discussion

## 4.1 Forward translation: relationship between facies and petrophysical properties

The forward translation results show systematic differences in performance across porosity, VCL, and permeability, reflecting the varying degree of coupling between facies and each petrophysical property (Figs. 4–6; Table 4).

The strong performance of the facies-to-porosity (Fig. 4) task is consistent with the direct geological relationship between depositional facies and porosity distribution. In clastic reservoirs, porosity is largely controlled by grain size, sorting, and compaction trends, all of which are embedded in facies architecture. As a result, the mapping between facies classes and porosity values is relatively smooth and laterally continuous, allowing the network to learn stable spatial patterns. This explains the high agreement observed in both image-based metrics and visual comparisons, particularly in the facies→porosity task (Fig. 4; Table 4).

A similar, although slightly weaker, behavior is observed for facies-to-VCL (Fig. 5) translation. VCL is directly linked to shale content and depositional energy, which are also facies-controlled properties. However, unlike porosity, VCL often exhibits sharper local transitions due to interbedding and mixed lithologies. These transitions introduce local variability that is less consistently captured by the model, leading to smoother boundaries in some generated slices. This explains the relatively high mean PA (class-balanced agreement) but lower overall PA and FWIoU compared to porosity (Fig. 5; Table 4).

Permeability shows the most challenging behavior among the forward tasks (Fig. 6). Unlike porosity and VCL, permeability is not solely controlled by facies but is also influenced by pore-throat geometry, connectivity, diagenesis, and sub-facies heterogeneity. These factors introduce non-linear and multi-scale variability that is not fully captured in a 2D image-to-image framework. As a result, the model reproduces the dominant permeability architecture but struggles to recover thin high-permeability streaks and sharp contrasts, leading to lower mIoU and more visible smoothing in local features. This explains both the lower mIoU reported in Table 4 and the visible smoothing in the challenging facies→permeability case (Fig. 6C1–C3).

## 4.2 Reverse translation: information content of petrophysical properties

The inverse translation results provide insight into how much facies information is contained in each petrophysical property (Fig. 7; Table 4).

VCL→facies achieved the highest mPA and mIoU among all tasks, supporting the visual separation observed in Figure 7B and the quantitative results in Table 4. This reflects the strong lithological signal carried by VCL, where shale-rich and clean-sand intervals are clearly distinguishable. The relatively distinct separation of these domains in the input space allows the model to recover facies classes more reliably and maintain sharper boundaries between them.

Porosity-to-facies also shows strong overall performance, particularly in PA and FWIoU, indicating good reconstruction of dominant facies architecture. However, porosity distributions often overlap across adjacent facies classes, especially in intermediate-quality sands. This overlap reduces class separability and leads to smoother transitions between predicted facies classes, even when the large-scale structure is correctly reproduced.

Permeability-to-facies remains the most ambiguous inverse problem. The weaker performance across all metrics reflects the indirect and multi-factor nature of permeability. Since permeability is influenced by factors beyond depositional facies, such as diagenetic alteration and pore connectivity, its spatial distribution does not uniquely define facies boundaries. Consequently, the model is able to recover first-order facies trends but shows increased confusion between neighboring classes and more fragmented boundaries.

### 4.3 Spatial continuity and multiscale behavior

The variogram-based validation provides an important perspective on model performance beyond pixel-level accuracy by comparing the reference and generated spatial-continuity behavior for porosity, permeability, facies, and VCL (Fig. 8). Across all domains, the close agreement between generated and reference variograms indicates that the model preserves the dominant spatial-continuity structure of the reservoir.

This result is particularly significant because it demonstrates that the model captures second-order spatial statistics, not only visual similarity. The agreement at short lags indicates preservation of local spatial variability, while agreement at longer lags reflects correct representation of large-scale geological structures.

The observed differences between properties can also be interpreted in terms of multiscale behavior. Porosity and VCL exhibit smoother spatial variability, making them more compatible with the convolutional structure of the network. In contrast, permeability contains stronger high-frequency components and sharper local contrasts, which are more difficult to reconstruct after image resizing and patch-based adversarial training. This explains both the higher variogram error and the visual smoothing observed in permeability-related tasks.

### 4.4 Implications for reservoir modeling workflows

The results demonstrate that the extended Pix2Geomodel workflow is capable of learning consistent cross-domain relationships even in a more complex reservoir setting with reduced vertical support. The model successfully reproduces the main geological architecture and spatial continuity across both forward and inverse translation tasks.

From a practical perspective, this indicates that image-to-image translation frameworks can serve as efficient tools for rapid geomodel generation and property prediction. The ability to perform bidirectional translation also provides flexibility, allowing both forward prediction (facies to properties) and inverse inference (properties to facies) within the same framework.

However, the results also highlight important limitations. The reduced accuracy in permeability-related tasks suggests that additional information or more advanced modeling strategies may be required to capture highly heterogeneous properties. Similarly, the smoothing of thin features indicates that higher-resolution training, multiscale approaches, or 3D extensions may further improve performance.

### 4.5 Comparison with the Original Pix2Geomodel Workflow and Practical Implementation Constraints

The present study extends the original Pix2Geomodel workflow from a feasibility demonstration to a more demanding robustness and transferability test. The previous implementation established that Pix2Pix-based image-to-image translation can learn relationships between facies and petrophysical properties in a more favorable reservoir setting. In contrast, the current study applies the workflow to a different and more geologically complex reservoir, with stronger facies heterogeneity, reduced vertical support, high-resolution image exports, and six bidirectional translation tasks involving facies, porosity, permeability, and VCL.

The core neural architecture remains the same as in the original Pix2Geomodel workflow: a U-Net generator, PatchGAN discriminator, adversarial loss, and L1 reconstruction loss. Therefore, the contribution of the present study is not the introduction of a new architecture, but the evaluation of whether the existing Pix2Geomodel framework can remain stable and geologically meaningful under more constrained data conditions. This distinction is important because model transferability is a critical requirement for practical reservoir modeling workflows.

A major practical difference in the current implementation is the treatment of high-resolution reservoir images. The previous Pix2Geomodel implementation used a smaller image size and a more standard Pix2Pix preprocessing workflow, whereas the present implementation required resizing large paired images to 2048 × 1024 pixels, padding, approximately 12% upscaling, random cropping, and consistent paired augmentation. These modifications were necessary to preserve spatial correspondence between input and target domains while allowing the model to train efficiently on large geological images (Table 3; Table 5).

The current study also expands the task definition. Instead of focusing only on selected property translations, the workflow evaluates six forward and inverse mappings: facies→porosity, facies→permeability, facies→VCL, porosity→facies, permeability→facies, and VCL→facies. This broader task design allows the model to be assessed not only as a property-prediction tool, but also as an inverse facies-reconstruction framework. The results show that porosity and VCL are more strongly coupled to facies architecture, whereas permeability remains more difficult because it is controlled by pore connectivity, diagenesis, and sub-facies heterogeneity in addition to facies class.

Another practical consideration is computational cost. Compared with the original lower-resolution GeoPix implementation, the current workflow uses substantially larger paired images resized to 2048 × 1024 px, which increases GPU memory demand and training time. Therefore, applications to larger reservoirs should consider hardware availability, batch-size limitations, checkpoint storage, and the trade-off between spatial resolution and computational efficiency.

Several implementation constraints should be considered when applying the workflow to other reservoirs. First, paired image-to-image translation requires strict pixel-wise alignment between input and target domains throughout export, resizing, augmentation, and pairing. Second, augmented-sample leakage must be avoided by splitting data at the original-layer level before augmentation. Third, color-encoded continuous properties are useful for image translation but should be complemented by property-space validation if direct petrophysical accuracy is required. Fourth, permeability-related tasks may require additional conditioning variables or more advanced architectures because permeability is not uniquely determined by facies architecture alone.

Finally, in the present implementation, each translation task required approximately 30 min of training on a single NVIDIA RTX A5500 GPU with 24 GB VRAM, indicating that the workflow is computationally feasible at the tested scale but may require additional optimization for larger 3D or field-scale applications.

Overall, the comparison shows that the present study provides a practical extension of Pix2Geomodel rather than a replacement of the original framework. Its value lies in demonstrating transferability, documenting implementation constraints, and identifying where future architectural improvements are most needed.

Table 5. Comparison between the original Pix2Geomodel implementation and the current Pix2Geomodel extension.

| **Aspect** | **Previous Pix2Geomodel** | **Current study** | **Interpretation** |
|---|---|---|---|
| Main objective | Demonstrate feasibility of Pix2Pix-based reservoir translation | Test robustness and transferability on a harder reservoir | Current work is an extension/validation study |
| Dataset | Previous reservoir case | New complex reservoir dataset | Stronger transferability test |
| Geological complexity | Lower facies complexity | Seven reservoir-quality classes and stronger heterogeneity | Harder learning problem |
| Vertical support | Larger number of slices | Reduced retained-layer support | More constrained training condition |
| Image size in code | 256 × 256 | 2048 × 1024 | Major practical workflow adaptation |
| Architecture | U-Net + PatchGAN Pix2Pix | U-Net + PatchGAN Pix2Pix | Architecture unchanged |
| Loss | Adversarial + L1 | Adversarial + L1 | Objective unchanged |
| Tasks | Selected facies/property mappings | Six bidirectional facies–property mappings | Broader evaluation |
| Evaluation | Visual outputs and standard metrics | Visual cases, image-space metrics, variograms | Stronger geological validation |

| Main limitation | Case-specific feasibility | 2D learning, RGB encoding, permeability complexity | Guides future development |
|---|---|---|---|
| Novel contribution | Demonstrated feasibility | Demonstrated transferability and constraints | Current novelty is practical and geological |
| Time and resources required | Single NVIDIA RTX A5500 GPU, 24 GB VRAM, approximately 1.4 hr per task. | Single NVIDIA RTX A5500 GPU, 24 GB VRAM, approximately 30 min per task. | Computationally feasible at the tested scale |

### 4.6 Limitations and future work

Building on the implementation constraints summarized in Section 4.5, this section focuses on the remaining methodological limitations and future development pathways. Despite the overall performance of the workflow, several limitations are observed, particularly at finer spatial scales and in properties with high variability.

First, the use of two-dimensional slice-based training limits the model's ability to fully capture three-dimensional geological continuity. While lateral spatial patterns are well preserved, vertical relationships between layers are not explicitly learned. This constraint may affect the representation of vertically connected features and layered heterogeneity.

Second, the preprocessing step involves resizing high-resolution images to a fixed training resolution, which can lead to loss of fine-scale details. This effect is most evident in thin features, sharp transitions, and narrow high-contrast zones, where the generated outputs show smoothing or partial attenuation. Properties with strong small-scale variability, such as permeability, are more sensitive to this limitation.

Third, the current framework relies on color-encoded image representations of continuous petrophysical properties. Although effective for image-to-image translation, this approach does not directly preserve physical units during training, which may limit direct quantitative interpretation beyond relative spatial patterns.

Future work can address these limitations through several extensions. Incorporating multiscale or edge-aware loss functions may improve the reconstruction of sharp boundaries and thin geological features. Higher-resolution training strategies, including tiled or patch-based approaches, could enhance the representation of small-scale heterogeneity. Extending the framework to three-dimensional architectures or incorporating multi-slice conditioning would allow explicit modeling of vertical continuity.

These limitations are consistent with recent developments in generative geomodeling. Diffusion-based models have shown improved fidelity, diversity, and conditioning consistency for reservoir facies generation, while statistical evaluation studies emphasize that generated geomodels should be assessed beyond visual similarity using spatial continuity, conditioning accuracy, and distributional agreement (Lee et al., 2025; Ovanger et al., 2025; Wang & Zhang, 2025). In parallel, 3D generative modeling and deep-learning-versus-geostatistics comparisons show that learned generative workflows can produce geologically realistic realizations efficiently, although their performance depends strongly on conditioning density, architecture, and validation strategy (Dong et al., 2026; Dupont & Bailey, 2018; Song & Wang; Lee et al., 2025; Liu et al., 2026; Ovanger et al., 2025; Wang et al., 2025). These studies support the need for future Pix2Geomodel extensions that incorporate multiscale learning, explicit spatial-continuity constraints, and 2.5D or 3D conditioning.

Based on the observed limitations, Figure 9 summarizes possible future extensions of the workflow. This figure is intended as a conceptual roadmap rather than an experimental comparison; the present study evaluates only the baseline Pix2Pix-based Pix2Geomodel framework.

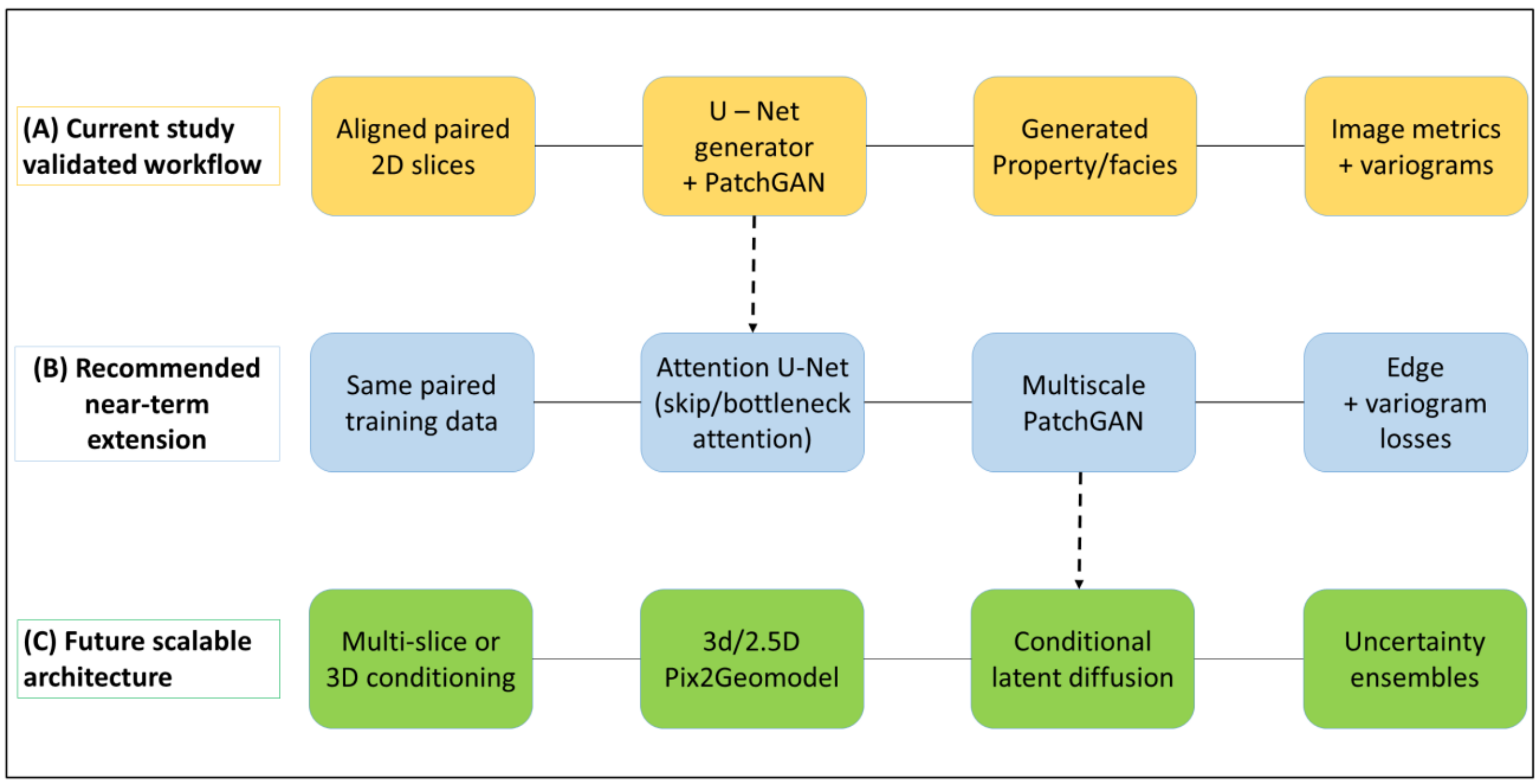

**Figure 9.** Conceptual roadmap for future Pix2Geomodel extensions. The current study evaluates the baseline Pix2Pix workflow using aligned paired 2D reservoir slices, a U-Net generator, a PatchGAN discriminator, and image-space plus variogram-based validation. The attention-enhanced, multiscale, 2.5D/3D, and diffusion-based modules shown in the figure are not part of the experiments reported in this study; they are presented as future development pathways motivated by the observed limitations in boundary preservation, fine-scale heterogeneity, vertical continuity, and uncertainty representation.

In addition, integrating physical-unit-based training or post-processing could strengthen the link between generated outputs and reservoir properties, enabling more direct quantitative validation. These developments would extend the applicability of the current workflow while building on its demonstrated capability for data-efficient and transferable facies–property translation.

## 5. Conclusions

This study evaluated the robustness and transferability of the Pix2Geomodel workflow for bidirectional facies–property translation in a more geologically complex reservoir setting with reduced vertical support. Using paired 2D slices extracted from facies, porosity, permeability, and VCL domains, six forward and inverse translation tasks were evaluated with image-space metrics, visual case analysis, and variogram-based spatial-continuity validation.

The results show that the workflow preserves the dominant geological architecture and main facies–property relationships across the evaluated tasks. The strongest performance was observed for facies–porosity and VCL–facies translation, reflecting the stronger geological coupling of porosity and clay volume with facies architecture. Permeability-related tasks were more challenging because permeability is controlled not only by facies, but also by pore connectivity, diagenesis, and sub-facies heterogeneity.

Variogram-based validation confirms that the generated outputs preserve the main spatial-continuity trends of the reference model, indicating that the workflow captures geological structure beyond pixel-level similarity. These results demonstrate that Pix2Geomodel can be transferred beyond its original case study and applied as a practical, data-efficient framework for rapid facies–property translation in complex reservoir settings.

The study also identifies important implementation constraints. Successful application requires strict paired-domain alignment, leakage-safe data splitting, careful treatment of color-encoded properties, and recognition of the limits of 2D slice-based learning. Future developments should focus on physical-unit-based training or validation, multiscale and edge-aware losses, attention-enhanced generators, 2.5D/3D conditioning, and diffusion-based uncertainty modeling. These extensions should be treated as subsequent developments rather than part of the current evaluated workflow.

### Acknowledgements

The authors acknowledge the Center for Integrative Petroleum Research and College of Petroleum Engineering and Geosciences at King Fahd University of Petroleum and Minerals for research support and computational resources.

### Conflicts of interest

The authors declare no competing interests.

### Data availability

The source code and example workflow for Pix2Geomodel are available at: https://github.com/ARhaman/GeoPix. Additional data and trained models are available from the corresponding author upon reasonable request, subject to data-sharing restrictions.